\documentclass{article}
\usepackage[dvips]{graphicx}

\newtheorem{thm}{Theorem}[section]

\newtheorem{cor}[thm]{Corollary}

\title{A note on the discrete-time evolutions of quantum walk on a graph}
\author{
{\small Yusuke HIGUCHI} \\ 
{\scriptsize Mathematics Laboratories, College of Arts and Sciences, Showa University}\\
{\scriptsize 4562 Kamiyoshida, Fujiyoshida,
Yamanashi 403-0005, Japan}\\ 
{\small Norio KONNO}\\
{\scriptsize Department of Applied Mathematics, 
Faculty of Engineering, Yokohama National University}\\
{\scriptsize Hodogaya, Yokohama 240-8501, Japan}\\
{\small Iwao SATO} \\ 
{\scriptsize Oyama National College of Technology} \\ 
{\scriptsize Oyama, Tochigi 323-0806, Japan} \\
{\small Etsuo SEGAWA} \\
{\scriptsize Graduate School of Information Sciences, Tohoku University} \\
{\scriptsize Sendai 980-8579, Japan}.}
\begin{document}
\maketitle

\begin{abstract}
For a quantum walk on a graph, there exist many kinds of operators 
for the discrete-time evolution. 
We give a general relation between the characteristic polynomial of 
the evolution matrix of a quantum walk on edges 
and that of a kind of transition matrix of a classical random walk on vertices. 
Furthermore we determine the structure of 
the positive support of the cube of some evolution matrix, 
which is said to be useful for isospectral problem in graphs,  
under a certain condition. 
\end{abstract}

{\bf keywords}: 
quantum walk, 
evolution matrix, characteristic polynomial, isospectral problem


\section{Introduction}
Recently many researchers in various fields pay attention to the quantum walk 
on graphs. 
Outstanding reviews are found, for example, in 
\cite{Ambainis2003, Kempe2003, Kendon2007,Konno2008b,VA}. 
Roughly speaking, 
a classical random walk on a graph presents the particle on some vertex moves 
to its neighbour one with some probability in one unit time, 
whereas a discrete-time quantum walk presents the quantum wave on some oriented 
edge travels to its neighbour one with some rate for its amplitude. 
In this note, we say {\it transition\/} or {\it adjacency\/} for a matrix 
giving the hopping rate between two vertices; {\it evolution\/} for 
a matrix giving the hopping rate between two oriented edges. 

Until now, 
the spectrum of the Grover evolution matrix 
${\bf U} = {\bf U}(G)$ of a regular graph $G$ is expressed 
in terms of that of the adjacency matrix ${\bf A}={\bf A}(G)$ 
of $G$ (cf. \cite{EmmsETAL2006,GG2010,KS2011}); moreover 
the spectra of the positive support ${\bf U}^{+}(G)$ of ${\bf U} (G)$ 
and the positive support $({\bf U}^{2})^{+}(G)$ of its square 
of a regular graph $G$ are also expressed in terms of 
that of ${\bf A} (G)$. 
On the other hand, a mapping property from the spectrum of 
the transition operator of a random walk on $G$ to that of 
the Szegedy evolution operator of a quantum walk, 
which is introduced firstly in \cite{Sz}, 
 is shown in \cite{Segawa}. 
One of our main purposes in this note is to give a generalized 
formula of the above. 

Let us explain our setting. 
Graphs treated here are finite only. 
Let $G=(V(G),E(G))$ be a connected graph 
(having possibly multiple edges and self-loops) 
with the set $V(G)$ of vertices and the set $E(G)$ of unoriented edges. 
We say two vertices $u$ and $v$ are {\it adjacent\/} if 
there exists an unoriented edge joining $u$ and $v$; $uv\in E(G)$. 
Considering each edge in $E(G)$ to have two orientations, 
we can introduce the set of all oriented edges; 
we denote it by $D(G)$. 
For an oriented edge $e\in D(G)$, the origin vertex and the terminal one 
of $e$ are denoted by $o(e)$ and $t(e)$, respectively;  
the inverse edge of $e^{-1}$ is denoted by $e$.
The {\em degree} $\deg v = \deg_{G}v$ of a vertex $v$ of $G$ stands for 
the number of oriented edges whose origin is $v$. 
Throughout this note, 
a connected graph $G$ is often assumed to have 
$n$ vertices and $m$ unoriented edges, 
$V(G)= \{ v_{1} , \ldots , v_{n} \} $ and 
$D(G)= \{ e_{1} , \ldots , e_{m} , 
e^{-1}_{1} , \ldots , e^{-1}_{m} \} $. 
Now let us give a weight $w$ on $D(G)$ such that $w(e)$ is a nonzero 
complex number for each $e\in D(G)$. 
With respect to this weight, we introduce three weighted matrices 
${\bf S}_{G}^{w}={\bf S}^{w}$, 
${\bf D}_{G}^{w}={\bf D}^{w}$ and ${\bf U}_{G}^{w}={\bf U}^{w}$. 
Firstly ${\bf S}^{w}$ is a weighted {\it transition\/} 
(or {\it adjacency\/}) $n\times n$ matrix, whose $(u,v)$-element stands for  
the hopping rate of particle's moving from $u\in V(G)$ to $v\in V(G)$, 
defined as follows: 
\begin{equation}\label{DefS}
({{\bf S}^{w}})_{u,v} =\left\{
\begin{array}{cl}
{\displaystyle \sum_{o(e)=u,t(e)=v}w^{*}(e) w(e^{-1})},  
& \mbox{if $uv\in E(G)$,} \\
0, & \mbox{otherwise,}
\end{array}
\right.
\end{equation}
where $w^{*}(e)$ is the complex conjugate of a complex number $w(e)$. 
Secondly ${\bf D}^{w}$ is a weighted {\it degree\/}  
$n\times n$ diagonal matrix, whose $(u,u)$-element stands for  
the weighted degree of $u\in V(G)$, 
defined as follows: 
\begin{equation}\label{DefD}
({\bf D}^{w})_{u,v} = \delta_{u,v}\cdot \sum_{o(e)=v}|w(e)|^{2}.  
\end{equation}
Lastly, for an arbitrary fixed real number $s$, 
${\bf U}^{w,s}$ is a weighted {\it evolution\/} 
 $2m\times 2m$ matrix, whose $(e,f)$-element stands for  
the hopping rate of wave's traveling from $e\in D(G)$ to $f\in D(G)$, 
defined as follows: 
\begin{equation}\label{DefU}
({\bf U}^{w,s})_{e,f}=\left\{
\begin{array}{cl}
s\cdot w(e) w^{*}(f^{-1})-\delta_{e^{-1},f},  & \mbox{if $o(e)=t(f)$,}\\
0, & \mbox{otherwise.}
\end{array}
\right.
\end{equation}
Here $\delta_{a,b}$ is the Kronecker delta function, that is, 
\[
\delta_{a,b}=\left\{
\begin{array}{ll}
1, & \mbox{if $a=b$,} \\
0, & \mbox{otherwise.}
\end{array}
\right.
\]
${\bf U}^{w,s}$ is a utility operator in the following sense. 
Depending on the choice of a weight $w$ and a real number $s$, 
the matrix ${\bf U}^{w,s}$ may be an evolution operator of a 
quantum walk  or what is called an edge matrix, 
which are discussed in Examples~\ref{ex1} or~\ref{ex2} in Section~2, 
respectively. 

Our first theorem is a kind of spectral mapping property between ${\bf S}^{w}$ 
and ${\bf U}^{w,s}$ as follows: 

\begin{thm}\label{thm1}
For any finite graph $G$, any weight $w$ and any real number $s$, 
we have
\[
\det (\lambda{\bf I}_{2m}-{\bf U}_{G}^{w,s}) 
=(\lambda^{2}-1)^{m-n}\det\left(
(\lambda^{2}-1){\bf I}_{n}-s\cdot\lambda {\bf S}_{G}^{w} + s\cdot {\bf D}_{G}^{w}
\right).
\]
\end{thm}

Setting some suitable weight $w$ and integer $s$, we can easily obtain 
from Theorem~\ref{thm1} 
all the previous results concerning spectra of quantum walks 
found in \cite{EmmsETAL2006,GG2010, KS2011,Segawa}. 
Details will be discussed in Section~2. 

As a branch of discrete spectral geometry, 
it is natural to ask what geometric property of graphs effects on  
the spectral structure of an evolution operator of a quantum walk. 
Here we shall focus on an isospectral problem, which is 
one of the actual and classical ones, in graph settings, 
to answer the question raised by M.~Kac \cite{Kac}: 
Can you hear the shape of a drum? 
More precisely, our interest is 
to find an evolution operator of a quantum walk 
and a wider class of graphs such that 
any pair of isospectral graphs in such a class are always isomorphic. 
For a type of {\it adjacency\/} matrix, 
which corresponds to a classical random walk, 
there are many kinds of 
construction for a pair of isospectral non-isomorphic graphs. 
Those can be seen, for instance, in \cite{Terras}. 
Recent years, for a type of {\it evolution\/} matrix for a quantum walk, 
research studies on isospectral problems are actively given in  
\cite{EmmsETAL2006,EmmsETAL2009,GambelETAL,GG2010,RenETAL,ShiauETAL}.
In those, important evolution matrices are as follows: 
the Grover matrix ${\bf U}$, its positive support ${\bf U}^{+}$, 
the positive support of its square $({\bf U}^{2})^{+}$ 
and that of the cube $({\bf U}^{3})^{+}$. 
Here the Grover matrix ${\bf U} ={\bf U} (G)=( U_{e,f} )_{e,f \in D(G)} $ 
of $G$ is defined by 
\begin{equation}\label{Grover}
U_{e,f} =\left\{
\begin{array}{ll}
2/\deg_{G}{o(e)}, & \mbox{if $t(f)=o(e)$ and $f \neq e^{-1} $, } \\
2/\deg_{G}{o(e)}-1, & \mbox{if $f= e^{-1} $, } \\
0, & \mbox{otherwise.}
\end{array}
\right. 
\end{equation}
Moreover the {\em positive support} $\>{\bf F}^+ =( F^+_{i,j} )$ of 
a real matrix ${\bf F} =( F_{i,j} )$ is defined as follows: 
\begin{equation}\label{PS}
F^+_{i,j} =\left\{
\begin{array}{ll}
1, & \mbox{if $F_{i,j} >0$, } \\
0, & \mbox{otherwise.}
\end{array}
\right.
\end{equation}
To review briefly the results in \cite{EmmsETAL2006} and so on, 
any pair of regular graphs isospectral for ${\bf U}$, ${\bf U}^{+}$ 
and $({\bf U}^{2})^{+}$ are also isospectral 
for the standard adjacency matrix ${\bf A}$; 
thus each of such matrices cannot distinguish two non-isomorphic graphs.   
All of those can be easily obtained from Theorem~\ref{thm1};  
details will be seen in Section~2.  
On the other hand, Emms et al. pointed out 
the property of $({\bf U}^{3})^{+}$, the positive support of ${\bf U}$ cubed, 
is entirely different from that of 
${\bf U}$, ${\bf U}^{+}$ and $({\bf U}^{2})^{+}$ in \cite{EmmsETAL2006}:   
for the known family of {\it strongly regular graphs\/} 
${\rm srg}(n,k,r,s)$ up to $n=64$,  
any two non-isomorphic graphs have been verified, with help of computers, 
to be non-isospectral for $({\bf U}^{3})^{+}$. 
Here a strongly regular graph  ${\rm srg}(n,k,r,s)$ with parameters $(n,k,r,s)$
is a $k$-regular on $n$ vertices such that 
any two adjacent vertices have exactly $r$ common neighbours 
and any two nonadjacent vertices have exactly $s$ common neighbours. 
We should remark that, any two graphs with same parameters are isospectral for 
the standard adjacency matrix ${\bf A}$, but are not always isomorphic. 
Refer to a standard text book, e.g. \cite{BH}. 
For example, it is known that the number of non-isomorphic graphs 
${\rm srg}(36,15,6,6)$ is 32,548; thus the result stated above says that 
all of these are naturally isospectral for ${\bf A}$, whereas 
any pair of those are non-isospectral for $({\bf U}^{3})^{+}$.
Also, in \cite{EmmsETAL2006}, 
two 4-regular non-isomorphic graphs on 14 vertices which are isospectral for 
$({\bf U}^{3})^{+}$ are stated, 
so the following interesting conjecture is proposed: 
For any strongly regular graphs with the same set of parameters, 
they are isospectral for $({\bf U}^{3})^{+}$ 
if and only if they are isomorphic. 

In this note, we shall give a kind of evidence that 
the structure of $({\bf U}^{3})^{+}$ is different from that of 
${\bf U}^{+}$ or ${\bf A}$ but not so far. 
Our second theorem is, for regular graphs with their girth greater than 4, 
to illustrate  
the difference between $({\bf U}^{3})^{+}$ and a polynomial of ${\bf U}^{+}$
in a simple form. Here the girth $g(G)$ of a graph $G$ is the length of 
a shortest cycle in $G$. 
Detail can be seen in Section~3. 

\begin{thm}\label{thm2}
Let $G$ be a connected $k$-regular graph such that 
$k\geq 3$ and its girth $g(G)\geq 5$. 
The positive support $({\bf U}^3 ) ^+ $ is of the form 
\[
( {\bf U}^{3} )^{+} =( {\bf U}^{+} )^{3} + {}^{T}{\bf U}^+,  
\]
where ${}^{T}{\bf U} $ stands for the transpose of ${\bf U}$. 
\end{thm}

The rest of this note is organized as follows. 
In Section~2, after giving the proof of Theorem~\ref{thm1}, 
we state, as some application, some characteristic polynomials and 
the spectra of evolution matrix and its positive support.
In Section~3, we treat the positive support of the cube of 
the Grover matrix and give the proof of Theorem~\ref{thm2}.

\section{Proof and application of Theorem~\ref{thm1}}
Suppose that 
$G$ is a connected graph with $n$ vertices and $m$ unoriented edges 
as is in Section~1. 
Let us first introduce a kind of coboundary operator ${\bf A}^{w}$
and a kind of shift operator ${\bf P}$ as follows:
${\bf A}^{w}$ is a $2m\times n$ complex valued matrix such that
\begin{equation}\label{DefAw}
({\bf A}^{w})_{e,v}=w(e^{-1})\cdot \delta_{t(e),v}
\end{equation}
and ${\bf P}$ is a $2m\times 2m$-matrix such that
\begin{equation}\label{DefP}
{\bf P}_{e,f}=\delta_{e^{-1},f}. 
\end{equation}
We write $({\bf A}^{w})^{*}$ for the adjoint matrix, which is called also 
the conjugate transpose matrix, of ${\bf A}$.  
It is easy to check that ${\bf P}^2={\bf I}_{2m}$ and that 
\begin{equation}
(({\bf A}^{w})^{*}){\bf A}^{w}= {\bf D}^{w},
\end{equation}
\begin{equation}
(({\bf A}^{w})^{*}){\bf P}{\bf A}^{w}= {\bf S}^{w}, 
\end{equation}
\begin{equation}
{\bf P}\left(s{\bf A}^{w}(({\bf A}^{w})^{*})-{\bf I}_{2m}\right)
      = {\bf U}^{w,s},\label{defU} 
\end{equation}
where $s$ is a fixed real number;  
${\bf D}^{w}$, ${\bf S}^{w}$ and ${\bf U}^{w,s}$ are defined as in Section~1. 

We shall give the proof of Theorem~\ref{thm1}. 

{\bf Proof of Theorem~\ref{thm1}}
It holds that 
\[
\begin{array}{rcl}
&& \det(\lambda{\bf I}_{2m}-{\bf U}^{w,s})=
\det\left(
\lambda{\bf I}_{2m}
       -{\bf P}\left(s{\bf A}^{w}(({\bf A}^{w})^{*})-{\bf I}_{2m}\right)
\right)\\
&=& \det\left(
\lambda{\bf I}_{2m}+{\bf P}-s{\bf P}{\bf A}^{w}(({\bf A}^{w})^{*})
\right)\\
&=& \det(\lambda{\bf I}_{2m}+{\bf P})
\det\left(
{\bf I}_{2m}-s{\bf P}{\bf A}^{w}(({\bf A}^{w})^{*})
(\lambda{\bf I}_{2m}+{\bf P})^{-1}
\right) 
\end{array}
\]
 for any generic $\lambda$. 
Here we should remark that 
\[
(\lambda{\bf I}_{2m}+{\bf P})(\lambda{\bf I}_{2m}-{\bf P})
=(\lambda^{2}-1){\bf I}_{2m}
\]
and that
\[
\begin{array}{rcl}
\det({\bf I}_{m}-{\bf K}{\bf L})&=&\det\left(
\left( 
\begin{array}{cc}
{\bf I}_{m}&-{\bf K}\\
{\bf 0}_{n,m}&{\bf I}_{n} 
\end{array}
\right)
\left( 
\begin{array}{cc}
{\bf I}_{m}&{\bf K}\\
{\bf L}&{\bf I}_{n} 
\end{array}
\right)
\right) \\ 
&=& 
\det \left(
\left( 
\begin{array}{cc}
{\bf I}_{m}&{\bf K}\\
{\bf L}&{\bf I}_{n} 
\end{array}
\right)
\left( 
\begin{array}{cc} 
{\bf I}_{m}&-{\bf K}\\
{\bf 0}_{n,m}& {\bf I}_{n}
\end{array}
\right)\right)\\
&=& \det({\bf I}_{n}-{\bf L}{\bf K}) 
\end{array}
\]
 for any $m\times n$-matrix ${\bf K}$ and 
any $n\times m$-matrix ${\bf L}$. 
Then we can see it holds that 
\[
\begin{array}{rcl}
&& \det\left(
{\bf I}_{2m}-s{\bf P}{\bf A}^{w}(({\bf A}^{w})^{*})
(\lambda{\bf I}_{2m}+{\bf P})^{-1}
\right)\\
&=& \det\left(
{\bf I}_{n}-(s/(\lambda^{2}-1))(({\bf A}^{w})^{*})(\lambda{\bf I}_{2m}-{\bf P})
{\bf P}{\bf A}^{w}
\right)\\
&=& \det\left(
{\bf I}_{n}-(s/(\lambda^{2}-1))(\lambda(({\bf A}^{w})^{*}){\bf P}{\bf A}^{w}-
(({\bf A}^{w})^{*}){\bf A}^{w})
\right).
\end{array}
\]
By (8) and (9), we obtain 
\[
\det(\lambda{\bf I}_{2m}-{\bf U}_{G}^{w,s}) 
=(\lambda^{2}-1)^{m-n}\det\left(
(\lambda^{2}-1){\bf I}_{n}-s\cdot\lambda {\bf S}_{G}^{w} + s\cdot {\bf D}_{G}^{w}
\right), 
\]
where both sides of the above are the polynomials of $\lambda$ of order $2m$. 
This completes the proof.
Q.E.D. 

{\bf Example 1}.
When $s=2$ and ${\bf D}^{w}={\bf I}_{n}$, that is,
\[
\sum_{o(e)=v} |w(e)|^{2}=1
\] 
for every vertex $v\in V(G)$, we can easily check that
${\bf U}^{w,2}$ becomes a unitary matrix. 
Thus it may be said that 
${\bf U}^{w,s}$ presents various types of 
evolution operators of quantum walks. 
Actually we denote by ${\bf C}_{v}$ a (local) unitary operator as follows:
for each vertex $v\in V(G)$, 
\begin{equation}
({{\bf C}_{v}})_{e,f} =\left\{
\begin{array}{cl}
2w(e) w^{*}(f)-\delta_{e,f},  
& \mbox{if $v=o(e)=o(f)$,} \\
0, & \mbox{otherwise.}
\end{array}
\right.
\end{equation}
Restricting the set of oriented edges to 
$D_{v}(G)=\{e;o(e)=v\}$, we can naturally identify ${\bf C}_{v}$ with 
$2{\bf w}_{v}{\bf w}_{v}^{*}-{\bf I}_{d_{v}}$,
where $d_{v}=\deg_{G}v$ and ${\bf w}_{v}$ is a column vector 
${\bf w}_{v}={}^{T}(w(e_{1}),\dots ,w(e_{d_{v}}))$ such that 
$e_{k}\in D_{v}(G)$ for each $k$. The expression above implies 
the {\it reflection\/} operator 
in the $d_{v}$-dimensional complex vector space. In this sense, 
${\bf C}_{v}$  is often called a local {\it quantum coin\/} at $v$ 
of reflection type. 
%
It should be noted that
\[
2{\bf A}^{w} (({\bf A}^{w})^{*}) -{\bf I}_{2m}=
{\bf P}(\oplus_{v\in V(G)}{\bf C}_{v}){\bf P}
\]
in (10); thus it holds that
\begin{equation}
{\bf U}^{w,2}= (\oplus_{v\in V(G)}{\bf C}_{v}){\bf P}.
\end{equation}
Hence the quantum walk induced by such a discrete time evolution ${\bf U}^{w,2}$
is so called a {\it coined quantum walk\/} 
(\cite{Ambainis2003,Kempe2003,Kendon2007}). 
We shall exhibit some illustrative examples below. 

Let $p:D(G)\to (0,1]$ be a transition probability such that 
\begin{equation}
\sum_{e:o(e)=v}p(e)=1,
\end{equation}
for every vertex $v\in V(G)$. 
A classical random walk on $G$ is defined by this probability $p$, that is,
a particle at $v=o(e)$ can be considered to move to a neighbour $t(e)$ 
along the oriented edge $e$ with probability $p(e)$ in one unit time. 
For a finite graph $G$, we consider the transition matrix ${\bf T}_{p}$ 
such that ${\bf T}_{p}$ is an $n\times n$-matrix and 
\begin{equation}
({\bf T}_{p})_{u,v}=\left\{
\begin{array}{cl}
{\displaystyle \sum_{o(e)=u,t(e)=v}p(e)},  & \mbox{if $uv\in E(G)$,} \\
0, & \mbox{otherwise.}
\end{array}
\right.
\end{equation}
With respect to the transition probability of a classical random walk, 
the evolution matrix of the Szegedy walk, which is a kind of quantum walk 
introduced in \cite{Sz}, 
is defined as follows (cf. \cite{Segawa,Sz}): ${\bf U}_{sz}$ is a 
$2m\times 2m$-matrix and 
\begin{equation}\label{Szegedy}
({\bf U}_{sz})_{e,f} =\left\{
\begin{array}{ll}
2\sqrt{p(e)p(f^{-1})}-\delta_{e^{-1},f}, & \mbox{if $t(f)=o(e)$, } \\
0, & \mbox{otherwise.}
\end{array}
\right. 
\end{equation}
Now let us set a weight $w$ as $w(e)=\sqrt{p(e)}$ and $s=2$ 
in Theorem~\ref{thm1}. 
Thus, by (1).(2),(3), 
 we obtain that ${\bf D}^{w}={\bf I}_{n}$, ${\bf U}^{w,s}={\bf U}_{sz}$ and 
\begin{equation}\label{DefS2}
({{\bf S}^{w}})_{u,v} =\left\{
\begin{array}{cl}
{\displaystyle \sum_{o(e)=u,t(e)=v}\sqrt{p(e)p(e^{-1})}},  & \mbox{if $uv\in E(G)$,} \\
0, & \mbox{otherwise.}
\end{array}
\right. 
\end{equation}
We denote ${\bf S}^{w}$ in (16) by ${\bf S}_{p}$ here. 
Thanks to Theorem~\ref{thm1}, we obtain the following formula, which recovers 
the result for finite graphs in \cite{Segawa}: 

\begin{cor}\label{Cor2.1}(cf.\cite{Segawa})
For the Szegedy matrix ${\bf U}_{sz}$ of $G$, we have
\[
\det ( \lambda {\bf I}_{2m} - {\bf U}_{sz} )= 
( \lambda^{2} -1)^{m-n} \det (( \lambda^2 +1) {\bf I}_{n} -2 \lambda {\bf S}_{p}). 
\]
\end{cor}

For a transition probability $p$, if there exists a positive valued function
 $m:V(G)\to (0,\infty )$ such that
\begin{equation}
m(o(e))p(e)=m(t(e))p(e^{-1})
\end{equation}
for every oriented edge $e\in D(G)$, $p$ is said to be {\it reversible\/}; 
the function $m$ is said to be a reversible measure for $p$ or 
for the random walk, which is unique, if exists, up to a multiple constant.  
If $p$ is reversible, 
it is easy to check that $M{\bf T}_{p}M^{-1}={\bf S}_{p}$, where 
$(M)_{u,v}=\sqrt{m(u)}\cdot\delta_{u,v}$; 
hence ${\bf T}_{p}$ and ${\bf S}_{p}$ are isospectral. 
As a representative examples of a reversible random walk, 
we may display the {\it simple\/} random walk on $G$,  
which is induced by $p$ such that $p(e)=1/\deg_{G}o(e)$ for every $e\in D(G)$. 
Obviously $m(u)=\deg_{G}u$ is a reversible measure for such $p$. 
We denote the transition matrix for the simple random walk by ${\bf T}_{0}$. 
The Szegedy matrix with respect to 
the simple random walk is called the Grover matrix, 
whose original form can be seen in \cite{Wat}. 
In fact, setting $p(e)=1/\deg_{G}o(e)$ in (15), we can 
get (4) introduced in Section~1. 
For the simple random walk ${\bf T}_{0}$, 
the standard adjacency matrix ${\bf A}$ can be expressed as 
\begin{equation}
{\bf A}={\bf D}{\bf T}_{0},
\end{equation}
where ${\bf D}$ is the standard degree matrix such that
\begin{equation}\label{DefSTD}
({\bf D})_{u,v}={\deg_{G}u}\cdot\delta_{u,v}.
\end{equation}
Combining the above with Corollary~\ref{Cor2.1}, we have also the 
following formula, which recovers the results seen 
in \cite{EmmsETAL2006, KS2011, Segawa}: 
\begin{cor}(cf.\cite{EmmsETAL2006, KS2011, Segawa})
For the Szegedy matrix ${\bf U}_{sz}$ with respect to a reversible random walk
 ${\bf T}_{p}$, we have
\[
\det ( \lambda {\bf I}_{2m} - {\bf U}_{sz} )= 
( \lambda^{2} -1)^{m-n} 
 \det (( \lambda^2 +1) {\bf I}_{n} -2 \lambda {\bf T}_{p}). 
\]
In addition, 
for the Grover matrix ${\bf U}$, we can express the above 
in terms of ${\bf A}$ as  
\[
\begin{array}{rcl}
\det ( \lambda {\bf I}_{2m} - {\bf U})
= \frac{( \lambda^{2} -1)^{m-n} \det (( \lambda^{2} +1) {\bf D} -2 \lambda {\bf A})}
{\prod_{v\in V(G)}\deg_{G}v}. 
\end{array}
\]
\end{cor}

{\bf Example 2}. 
Here let us set a weight $w$ as $w(e)=1$ for any oriented edge $e\in D(G)$ 
and $s=1$; for such $w$ and $s$, we denote 
${\bf S}^{w}$, ${\bf D}^{w}$ and ${\bf U}^{w,s}$ by 
${\bf S}^{1}$, ${\bf D}^{1}$ and ${\bf U}^{1,1}$, respectively. 
Thus, by (1),(2),(3), 
 we obtain the following: ${\bf S}^{1}$ becomes 
the standard adjacency matrix ${\bf A}$, that is, 
\begin{equation}\label{DefAdj}
({{\bf A}})_{u,v} =\left\{
\begin{array}{cl}
{\displaystyle \sum_{o(e)=u,t(e)=v}1},  & \mbox{if $uv\in E(G)$,} \\
0, & \mbox{otherwise;}
\end{array}
\right. 
\end{equation}
${\bf D}^{1}$ becomes 
the standard degree matrix ${\bf D}$ as is seen in (19);  
${\bf U}^{1,1}$ becomes a $2m\times 2m$-matrix such that 
\begin{equation}\label{U11}
({\bf U}^{1,1})_{e,f} =\left\{
\begin{array}{ll}
1, & \mbox{if $o(e)=t(f)$ and $f\not= e^{-1}$, } \\
0, & \mbox{otherwise.}
\end{array}
\right.
\end{equation}
For any graph $G$ such that $\min_{v\in V(G)}\deg_{G}v\geq 2$, 
we can easily see that the positive support ${\bf U}^{+}$ of the 
the Grover matrix ${\bf U}$ introduced in Section~1 coincides with 
${\bf U}^{1,1}$ in (21). 
In the context of the Ihara zeta function of a graph
(see \cite{Bass1992,Hashimoto1989,Ihara1966,KS2011}), 
the concept of {\it edge matrix\/} plays an important role.  
For a $2m \times 2m$ matrix
${\bf B} = {\bf B} (G)=( {\bf B}_{e,f} )_{e,f \in D(G)} $ such that
\begin{equation}
{\bf B}_{e,f} =\left\{
\begin{array}{ll}
1, & \mbox{if $t(e)=o(f)$, } \\
0, & \mbox{otherwise,}
\end{array}
\right.
\end{equation}
the {\em edge matrix} of $G$ is defined as ${\bf B} - {\bf P}$, 
which obviously coincides with ${}^{T}{\bf U}^{1,1}$. 
As is also shown in \cite{RenETAL}, we have
\begin{equation}\label{edM}
{\bf B} - {\bf P} ={}^T {\bf U}^{+}
\end{equation}
for any graph $G$ such that $\min_{v\in V(G)}\deg_{G}v\geq 2$. 
Summarizing the above with Theorem~\ref{thm1}, 
we obtain the following formula, which recovers 
the results in \cite{EmmsETAL2006,GG2010,KS2011}:
\begin{cor}\label{Cor2.3}(cf.\cite{EmmsETAL2006,GG2010,KS2011})
For a graph $G$ such that $\min_{v\in V(G)}\deg_{G}v\geq 2$ and  
the positive support ${\bf U}^{+}$ of 
the Grover matrix ${\bf U}$, we have
\[
\det \left( \lambda {\bf I}_{2m} - {\bf U}^+ \right) 
=( \lambda {}^2 -1)^{m-n} \det \left( ( \lambda {}^2 -1) {\bf I}_n 
- \lambda {\bf A} + {\bf D} \right). 
\]
In addition, if $G$ is a connected $k$-regular graph with $k \geq 2$, 
${\bf U}^{+} $ has $2n$ eigenvalues of the form 
\[
\lambda = \frac{\lambda {}_{A} }{2} \pm 
\sqrt{-1}\sqrt{k-1- \lambda {}^{2}_{A} /4} , 
\]
where $\lambda {}_A $ is an eigenvalue of the matrix ${\bf A}$. 
The remaining $2(m-n)$ eigenvalues of ${\bf U}^+$ are $\pm 1$ with equal multiplicities. 
\end{cor}

On the positive support $( {\bf U}^2 )^+$ 
of the Grover matrix ${\bf U}$ squared for a regular graph, 
its eigenvalues are expressed by those of ${\bf A}$ in \cite{EmmsETAL2006}; 
another proof by using different methods is also given in \cite{GG2010}. 
\begin{thm}[\cite{EmmsETAL2006}]\label{Emmsthm}
Let $G$ be a connected $k$-regular graph with $n$ vertices and $m$ edges. 
Suppose that $k\geq 2$. 
The positive support $({\bf U}^2 ) ^+ $ has $2n$ eigenvalues of the form 
\[
\lambda = \frac{\lambda_{A}^{2} -2k+4}{2} 
\pm\sqrt{-1}
\lambda_{A} \sqrt{k-1- \lambda ^{2}_{A}/4 }
\]
The remaining $2(m-n)$ eigenvalues of ${\bf U}^+$ are $2$. 
\end{thm}

Let us close this section with giving still another proof of 
Theorem~\ref{Emmsthm} in virtue of expressing
the characteristic polynomial of $( {\bf U}^2 )^{+}$ 
in terms of ${\bf A}$ directly. 

{\bf Proof of Theorem~\ref{Emmsthm} }
It is easy to see that 
\begin{equation}
( {\bf U}^2 )^+ =({\bf U}^+ )^2+{\bf I}_{2m}  
\end{equation}
for $k\geq2$ (cf.\cite{GG2010}) , so we have  
\[
\begin{array}{rcl}
\det ( \lambda {\bf I}_{2m} - ({\bf U}^{2})^{+} )
&=& \det ( \lambda {\bf I}_{2m} - (({\bf U}^+ )^2+{\bf I}_{2m} ) ) \\
&=& \det ((\lambda -1) {\bf I}_{2m} - ({\bf U}^+ )^2 ) . 
\end{array} 
\]

Moreover it follows from Corollary~\ref{Cor2.3} that 
\[
\det ( \lambda {\bf I}_{2m} - ({\bf U})^{+} )= 
(\lambda^{2} -1)^{m-n} \cdot \det ((\lambda^{2} +k-1) 
  {\bf I}_{2m} - \lambda {\bf A}), 
\]
where ${\bf D}= k {\bf I}_n $ here since $G$ is $k$-regular.
Now let us denote $\det (\lambda {\bf I} - {\bf M} )$ 
by $ \varphi (\lambda ; {\bf M} )$ for a square matrix ${\bf M} $. 
Then it holds that  
$ \varphi (\mu ; {\bf U}^+ )= \det (\mu {\bf I}_{2m} - {\bf U}^+ )$ and 
$ \varphi (- \mu ; {\bf U}^+ )=(-1 )^{2m} \det (\mu {\bf I}_{2m} + {\bf U}^+ )$. 
So we have 
\[
\begin{array}{rcl}
\varphi (\mu ; {\bf U}^+ ) \varphi (- \mu ; {\bf U}^+ )
&=&\det (\mu {\bf I}_{2m} - {\bf U}^+ ) \det (\mu {\bf I}_{2m} + {\bf U}^+ ) \\
&=& \det (\mu {}^2 {\bf I}_{2m} -( {\bf U}^+ )^2 ) . 
\end{array}
\]
Simultaneously we have 
\[
\varphi (\mu ; {\bf U}^+ ) \varphi (- \mu ; {\bf U}^+ )=
( \mu {}^2 -1)^{2m-2n} \cdot 
\det \left((\mu {}^2 +k-1)^2 {\bf I}_{2m} - \mu {}^2 {\bf A}^2\right). 
\]
Here putting $\mu^2 = \lambda -1$, we get 
\begin{equation}
\begin{array}{rcl}
&&\det ( \lambda {\bf I}_{2m} - ({\bf U}^2 )^+ )
= \det ((\lambda -1) {\bf I}_{2m} - ({\bf U}^+ )^2 ) \\
&=&( \lambda -2 )^{2m-2n} \cdot 
\det (( \lambda +k-2)^2 {\bf I}_{2m} -( \lambda -1) {\bf A}^2) . 
\end{array}
\end{equation}
Therefore, it follows that 
\[
\det ( \lambda {\bf I}_{2m} -( {\bf U}^2 )^+ ) 
=( \lambda -2)^{2m-2n} 
\prod_{ \lambda {}_A \in Spec ({\bf A})} 
(\lambda^{2} +(2k-4- \lambda^{2}_{A} ) \lambda +(k-2 )^{2} + \lambda^{2}_{A} ). 
\]
Solving $\lambda {}^2 +(2k-4- \lambda {}^2_A ) \lambda +(k-2 )^2 + \lambda {}^2_A = 0$, we can get the result. 
Q.E.D. 

As a conclusion in this section, we may state the following: 
if two $k$-regular graphs which are isospectral for ${\bf A}$, 
then they are also isospectral 
for ${\bf U}$, ${\bf U}^{+}$ and $({\bf U}^{2})^{+}$; 
thus each of such matrices cannot distinguish two non-isomorphic graphs.     

\section{The positive support of the cube of the Grover matrix of a graph}
Let $G$ be a connected graph. 
Then a {\em path $P$ of length $\ell$} in $G$ is defined as 
a sequence $P=(e_1, \ldots ,e_{\ell} )$ of $\ell$ 
oriented edges in $D(G)$ such that  
$t( e_i )=o( e_{i+1})$ $ (1 \leq i \leq \ell-1)$.  
We may write 
$P=(v_0, e_1 , v_1, \cdots ,v_{\ell-1}, e_{\ell} , v_{\ell} )$, 
if $o(e_i)=v_{i-1} $ and $t(e_i )= v_i $ for $i=1, \ldots, \ell$, 
The path $P$ is called a {\em cycle} if $v_{0}=v_{\ell}$. 
In addition, 
a cycle $C=(v_0, e_1 , v_1, \cdots ,v_{\ell-1}, e_{\ell} , v_{0} )$ 
is called {\em essential} if $e_{1}^{-1}\not=e_{\ell}$ and 
all the vertices of $C$ are mutually distinct. 
The {\em girth} $g(G)$ of a graph $G$ is defined as 
the minimum length of essential cycles in $G$. 

Assuming that 
$G$ is a connected $k$-regular graph with $k\geq 3$ and $g(G)\geq 5$,  
we shall give the proof of Theorem~\ref{thm2} as is seen in Section~1. 

{\bf Proof of Theorem~\ref{thm2}}
Here $G$ is a connected $k$-regular graph ($k\geq 3$) 
with $n$ vertices and $m$ edges. 
As is in Example~2, we shall put $w$ as $w(e)=1$ 
for any oriented edge $e\in D(G)$ and $s=1$. 
Let us denote ${\bf A}^{w}$ in (6) by ${}^{T}{\bf D}_{h}$ 
and ${\bf D}_{h}{\bf P}$ by  ${\bf D}_{t}$, where ${\bf P}$ is defined as 
in (7) in Section~2. Hence they can be expressed as 
\begin{equation}
( {\bf D}_{h} )_{v,e} =\left\{
\begin{array}{ll}
1, &\mbox{if $t(e)=v$, } \\
0, &\mbox{otherwise,}
\end{array}
\right. 
\end{equation}
\begin{equation}
( {\bf D}_{t} )_{v,e}  
=\left\{
\begin{array}{ll}
1, &\mbox{if $o(e)=v$, } \\
0, &\mbox{otherwise.}
\end{array}
\right.
\end{equation}
In addition, it follows from (1),(2),(3) and Example 2  
that  
\begin{equation}
{\bf D}_{h} ({}^T {\bf D}_{t}) = {}^T {\bf A} \  and \  
({}^T {\bf D}_{t}) {\bf D}_{h} = {}^T {\bf B}  
\end{equation}
and that 
\begin{equation}
{\bf U} = \frac{2}{k} ({}^T {\bf D}_{t}) {\bf D}_{h} - {\bf P}. 
\end{equation}
We shall consider the structure of 
the positive support $( {\bf U}^3 )^+$ of the cube of 
the Grover matrix ${\bf U} $.     
Since all nonzero elements of ${\bf B} $ and $ {}^T {\bf U} $ 
are in the same place, 
all nonzero elements of ${\bf B}^3 $ and $ {}^T {\bf U}^3 $ are 
in the same place; we treat ${\bf B}^3 $ and $ {}^T {\bf U}^3 $ in parallel. 
Let us denote here ${}^{T}{\bf U}^{+}= {\bf B} - {\bf P}$ 
in (23) by ${\bf Q}$. 
Thus we have 
\[
\begin{array}{rcl}
{\bf B}^3 &=& ( {\bf Q} + {\bf P} )^3 \\
&=& {\bf Q}^3 + {\bf Q}^2 {\bf P} + {\bf Q} {\bf P} {\bf Q} + {\bf P} {\bf Q}^2\\ 
&& \quad + {\bf Q} {\bf P}^2 
+ {\bf P}^2 {\bf Q} + {\bf P}  {\bf Q} {\bf P} + {\bf P}^3 . 
\end{array}
\]
Now we divide the relation of oriented edges $e$ and $f$ of 
the nonzero $(e,f)$-element of $( {}^T {\bf U} )^3 $ 
into the eight cases in Figure 1. 
In fact, the cases I, II, III, IV, V, VI, VII and VIII correspond to the matrices 
${\bf Q}^3 $, ${\bf Q}^2 {\bf P} $, $ {\bf Q} {\bf P} {\bf Q} $, ${\bf P} {\bf Q}^2 $, 
$ {\bf Q} {\bf P}^2 $, $ {\bf P}^2 {\bf Q} $, ${\bf P}  {\bf Q} {\bf P} $ and $ {\bf P}^3 $, 
respectively. 

For a path $P=(e_{1}, \cdots , e_{\ell})$ in $G$, 
we say that $P$ is an {\em $(e_{1} , e_{\ell} )$-path}; 
if $e^{-1}_{i+1} =e_{i}$ for some $i (1 \leq i \leq \ell-1)$, 
we say that a path $P=(e_{1}, \ldots ,e_{\ell} )$ has a {\em backtracking}. 
Let us count the number of backtrackings in an $(e,f)$-path in each case. 
In the case I, an $(e,f)$-path has no backtracking;  
in the cases II, III and IV, an $(e,f)$-path has exactly one backtracking; 
in the cases V, VI and VII, an $(e,f)$-path has exactly two backtrackings; 
in the case VIII, an $(e,f)$-path has exactly three backtrackings.  
Then we can see that 
the elements of ${}^T {\bf U}^3 $ corresponding to nonzero elements of 
${\bf Q}^2 {\bf P} $, $ {\bf Q} {\bf P} {\bf Q} $, ${\bf P} {\bf Q}^2 $ 
and $ {\bf P}^3 $ are negative. 
Furthermore, the elements of ${}^T {\bf U}^3 $ 
corresponding to nonzero elements of 
${\bf Q}^3 $, $ {\bf Q} {\bf P}^2 $, $ {\bf P}^2 {\bf Q} $ and 
${\bf P}  {\bf Q} {\bf P} $ are positive. 

If $t(e)=o(f)$, then nonzero $(e,f)$-elements of $ {\bf Q} {\bf P} {\bf Q} $, $ {\bf Q} {\bf P}^2 $ and 
$ {\bf P}^2 {\bf Q} $ are overlapped. 
Then we have 
\[ 
( {\bf U}^3 )_{e,f} = \frac{2}{k} ( \frac{2}{k} -1) \frac{2}{k} \cdot (k-2)
+2\cdot\frac{2}{k} ( \frac{2}{k} -1)^{2} =0. 
\]  
Thus all positive elements of $( {}^T {\bf U} )^3 $ and 
${\bf Q}^3 + {\bf P} {\bf Q} {\bf P}$ are in the same place;  
it holds that 
\begin{equation}\label{eq28}
( {}^T {\bf U}^3 )^+ =( {\bf Q}^3 + {\bf P}  {\bf Q} {\bf P} )^+ . 
\end{equation}
We first show that 
nonzero element of ${\bf Q}^3 $ and ${\bf P}  {\bf Q} {\bf P} $ 
are not overlapped. 
Let us assume 
that a nonzero $(e,f)$-element of ${\bf Q}^3 $ and 
${\bf P} {\bf Q} {\bf P}$ are overlapped:  
there exists an essential cycle of length $4$ from $e$ to $f$ in $G$, 
which contradicts $g(G)>4$. 
Next we show that  all nonzero elements of 
two matrices $ {\bf Q}^{3}$ and ${\bf P}{\bf Q}{\bf P}$ are $1$. 
It is trivial that all nonzero elements of ${\bf P}{\bf Q}{\bf P}$ are $1$. 
Then let us assume that 
an $(e,f)$-element of $ {\bf Q}^3 $ is not less than $2$:  
there exist two distinct $(e,f)$-paths 
$P=(e,g,h,f)$ and $Q=(e,g_1 , h_1, f)$ in $G$ and then   
the cycle $(g,h,h^{-1}_1 , g^{-1}_1 )$ is an essential cycle of length $4$
 in $G$, which contradicts the assumption $g(G)>4$. 

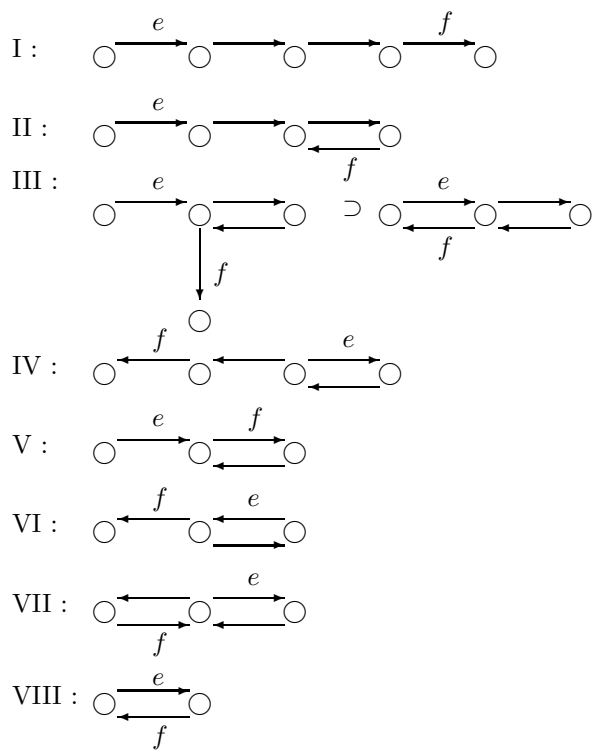
\begin{figure}
\begin{center}
\begin{picture}(200,270)
\put(9,250){I : }
\put(62,260){$e$}
\put(44,250){\circle{8}}
\put(48,255){\vector(1,0){27}}
\put(80,250){\circle{8}}
\put(85,255){\vector(1,0){27}}
\put(116,250){\circle{8}}
\put(121,255){\vector(1,0){27}}
\put(152,250){\circle{8}}
\put(157,255){\vector(1,0){27}}
\put(170,260){$f$}
\put(188,250){\circle{8}}
\put(9,220){II : }
\put(62,230){$e$}
\put(44,220){\circle{8}}
\put(48,225){\vector(1,0){27}}
\put(80,220){\circle{8}}
\put(85,225){\vector(1,0){27}}
\put(116,220){\circle{8}}
\put(121,225){\vector(1,0){27}}
\put(152,220){\circle{8}}
\put(148,215){\vector(-1,0){27}}
\put(134,205){$f$}
\put(9,200){III : }
\put(62,200){$e$}
\put(44,190){\circle{8}}
\put(48,195){\vector(1,0){27}}
\put(80,190){\circle{8}}
\put(85,195){\vector(1,0){27}}
\put(116,190){\circle{8}}
\put(112,185){\vector(-1,0){27}}
\put(80,185){\vector(0,-1){27}}
\put(80,150){\circle{8}}
\put(85,165){$f$}
\put(134,190){$\supset$}
\put(170,200){$e$}
\put(152,190){\circle{8}}
\put(157,195){\vector(1,0){27}}
\put(188,190){\circle{8}}
\put(184,185){\vector(-1,0){27}}
\put(170,175){$f$}
\put(193,195){\vector(1,0){27}}
\put(224,190){\circle{8}}
\put(220,185){\vector(-1,0){27}}
\put(9,130){IV : }
\put(62,140){$f$}
\put(44,130){\circle{8}}
\put(76,135){\vector(-1,0){27}}
\put(80,130){\circle{8}}
\put(112,135){\vector(-1,0){27}}
\put(116,130){\circle{8}}
\put(134,140){$e$}
\put(121,135){\vector(1,0){27}}
\put(152,130){\circle{8}}
\put(148,125){\vector(-1,0){27}}
\put(9,100){V : }
\put(62,110){$e$}
\put(44,100){\circle{8}}
\put(49,105){\vector(1,0){27}}
\put(80,100){\circle{8}}
\put(85,105){\vector(1,0){27}}
\put(116,100){\circle{8}}
\put(98,110){$f$}
\put(112,95){\vector(-1,0){27}}
\put(9,70){VI : }
\put(62,80){$f$}
\put(44,70){\circle{8}}
\put(76,75){\vector(-1,0){27}}
\put(80,70){\circle{8}}
\put(112,75){\vector(-1,0){27}}
\put(116,70){\circle{8}}
\put(98,80){$e$}
\put(85,65){\vector(1,0){27}}
\put(9,40){VII : }
\put(62,25){$f$}
\put(44,40){\circle{8}}
\put(49,35){\vector(1,0){27}}
\put(80,40){\circle{8}}
\put(76,45){\vector(-1,0){27}}
\put(85,45){\vector(1,0){27}}
\put(116,40){\circle{8}}
\put(98,50){$e$}
\put(112,35){\vector(-1,0){27}}
\put(9,5){VIII : }
\put(62,12){$e$}
\put(44,5){\circle{8}}
\put(49,10){\vector(1,0){27}}
\put(80,5){\circle{8}}
\put(76,0){\vector(-1,0){27}}
\put(62,-10){$f$} 
\end{picture}
\end{center}
\caption{\rm The nonzero $(e,f)$-array of $( {}^T {\bf U} )^3 $. }
\end{figure}
%
%
Thus the expression (30) becomes the following form: 
\[
( {\bf U}^3 )^+ =( {}^T {\bf Q} )^3 + {\bf P} {}^T {\bf Q} {\bf P} . 
\]
Since $ {}^T {\bf Q} = {}^T {\bf B} - {\bf P} = {\bf U}^+ $, 
we have 
\[
( {\bf U}^3 )^+ =( {\bf U}^+ )^3 + {\bf P} {\bf U}^+ {\bf P} . 
\] 
and 
\[
\begin{array}{rcl}
{\bf P} {\bf U}^+ {\bf P}&=&
{\bf P} ( {}^T {\bf D}_t {\bf D}_h - {\bf P} ) {\bf P} 
={\bf P}( {}^T {\bf D}_{t}){\bf D}_{h} {\bf P} - {\bf P}^3 \\ 
&=&{}^T {\bf D}_h {\bf D}_t - {\bf P} = {}^T {\bf U}^+ .  
\end{array}
\] 
Hence we obtain
\[
( {\bf U}^3 )^+ =( {\bf U}^+ )^3 + {}^T {\bf U}^+ . 
\]
Q.E.D.

%
%
%
%
%
%
%
%
\section*{Acknowledgments}
We would like to thank Doctor Krystal Guo for many valuable comments and 
many helpful suggestions. 
YuH's work was supported in part 
by JSPS Grant-in-Aid for Scientific Research (C)~20540113, 25400208 
and (B)~24340031. 
NK and IS also acknowledge financial supports of 
the Grant-in-Aid for Scientific Research (C) 
from Japan Society for the Promotion of Science 
(Grant No.~24540116 and No.~23540176, respectively).
ES thanks to the financial support of
the Grant-in-Aid for Young Scientists (B) of Japan Society for the
Promotion of Science (Grant No. 25800088). 
%
%
\begin{small}

\end{small}

\end{document}